\documentclass[letter,apj,numberedappendix,appendixfloats]{emulateapj-rtx4}

\usepackage[english]{babel}
\usepackage[utf8x]{inputenc}
\usepackage[T1]{fontenc}
\usepackage{color}
\usepackage{setspace}   

\usepackage{amsmath}
\usepackage{graphicx}

\begin{document}

\title{Stochastic Speckle Discrimination with Time-Tagged Photon Lists: Digging Below the Speckle Noise Floor}
\author{Alex B.~Walter\altaffilmark{1}, Clinton Bockstiegel\altaffilmark{1}, Timothy D.~Brandt\altaffilmark{1}, Benjamin A.~Mazin\altaffilmark{1}}
\altaffiltext{1}{Department of Physics, University of California, Santa Barbara, Santa Barbara, CA 93106, USA}


\begin{abstract}
We present an algorithm that uses the distribution of photon arrival times to distinguish speckles from incoherent sources, like planets and disks, in high contrast images. Using simulated data, we show that our approach can overcome the noise limit from fluctuating speckle intensity. The algorithm is likely to be most advantageous when a coronagraph limits the coherent diffraction pattern in the image plane but the intensity is still strongly modulated by fast-timescale uncorrected stellar light, for example from atmospheric turbulence. These conditions are common at small inner working angles and will allow probing of exoplanet populations at smaller angular separations. The technique requires a fast science camera that can temporally resolve the speckle fluctuations, and the detection of many photons per speckle decorrelation time. Since the algorithm directly extracts the incoherent light, standard differential imaging post-processing techniques can be performed afterwards to further boost the signal. 

\end{abstract}

\section{Introduction}

Direct imaging is a challenging exoplanet discovery and characterization technique due to the extreme contrast ($<$10$^{-4}$ for ground based targets) and small angular separations ($\lesssim$1$^{\prime\prime}$) between the planetary companion and its stellar host. Despite this, adaptive optics (AO) and coronagraphy have enabled the discovery of planets up to $\sim$10$^6$ times fainter than their host stars \citep{Marois+Macintosh+Barman+etal_2008,Lagrange+Bonnefoy+Chauvin+etal_2010,Kuzuhara+Tamura+Kudo+etal_2013,Macintosh+Graham+Barman+etal_2015,Keppler+Benisty+Muller+etal_2018}. Imaging an exoplanet requires subtracting the light of its host star in the form of the point-spread function (PSF). If this background were static and could be subtracted perfectly, exoplanet imaging would be limited only by the photon shot noise of the bright host star. Instead, high-contrast imaging is limited by uncontrolled scattered and diffracted light, which produces a coherent speckle halo in the image plane \citep{Guyon_2005}.   

Fast atmospheric speckles average down over an observation, while slower, quasistatic speckles must be removed using post-processing techniques. Angular differential imaging \citep[ADI,][]{Marois+Lafreniere+Doyon+etal_2006} exploits the rotation of the Earth, and hence the field-of-view of an altitude-azimuth telescope, to distinguish diffraction speckles from astrophysical sources. Spectral differential imaging \citep[SDI,][]{Racine+Walker+Nadeau+etal_1999,Marois+Doyon+Racine+etal_2000,Sparks+Ford_2002} uses the scaling of diffraction speckles with wavelength. Since the initial development of ADI and SDI, a variety of post-processing algorithms have refined their approaches to dig deeper into the stellar PSF \citep[e.g.][]{Lafreniere+Marois+Doyon+etal_2007,Soummer+Pueyo+Larkin_2012,Marois+Correia+Galicher+etal_2014}.

The time variability and chromaticity of quasi-static speckles limit the performance of ADI and SDI \citep{Gerard+Marois+Currie+etal_2019}. Both techniques also suffer at small separations where exoplanets are more likely to hide. The speckle spectral dispersion used by SDI is proportional to the separation: close to the star, it becomes smaller than the planet's PSF. For ADI, the arclength traced by the companion's sky rotation is proportional to the separation. Furthermore, the precision of the background estimate for PSF subtraction is limited by low counting statistics at small separations \citep{Mawet_2014}. Even without these issues, the variability induced by speckle fluctuations can dominate the photon noise and be well above the shot noise expected from the total number of photons. 

Stochastic Speckle Discrimination \citep[SSD,][]{Gladysz_2008} is a post-processing technique designed to reduce the additional noise caused by speckle fluctuations by temporally resolving them. It relies on the difference in photon arrival time statistics between a planet/extended source and the off-axis residual stellar speckles \citep{Canales1999, Cagigal_2001, aime04b, fitzgerald_2006, Soummer_2007}. With millisecond imaging cameras, SSD-like approaches have been shown to reduce speckle noise from even fast atmospheric speckles and improve the contrast limit for the detection of substellar companions \citep{Gladysz_2010, Frazin_2016, Meeker2018, Stangalini_2018}.

In this paper, we present an improved version of SSD to exploit noise-free photon-counting cameras like MEC, the Microwave Kinetic Inductance Detector (MKID) Exoplanet Camera \citep{Walter2018, Meeker2018} on Subaru Telescope's SCExAO instrument \citep{Lozi2018}. Our approach statistically distinguishes a combination of constant and speckle intensity (both ultimately from the bright star) from incoherent light (from a planet or disk). At small separations, where ADI and SDI are least effective and the scientific questions are most pressing \citep{Mawet2012}, we expect SSD to offer strong improvements in the limiting detection contrast. This paper describes the new photon counting SSD technique and demonstrates its performance on simulated data. 

We organize the paper as follows. In Section \ref{sec:sim} we detail the simulation of photon lists to emulate the data expected from photon-counting cameras like MEC. Section \ref{sec:binned} describes a formal extension of previous SSD analysis techniques with millisecond images using a maximum likelihood algorithm. We simulate the performance for various atmospheric conditions, planet brightnesses, and effective exposure times. Section \ref{sec:binfree} presents the new photon-counting SSD algorithm that estimates the incoherent light from a companion or disk directly from individual photon arrival timestamps. We demonstrate the algorithm on a simulated telescope image. We discuss the main results in Section \ref{sec:discuss}, and conclude with Section \ref{sec:conclusions}. 

\section{Simulating Photon Arrival Times} \label{sec:sim}

\subsection{Modeling the Stellar Speckle Intensity}
The statistics governing the off-axis intensity distribution of coherent light with a partially developed speckle pattern have been studied at length. Originally derived by \cite{Goodman1975} and verified experimentally by \cite{Cagigal_2001} and \cite{fitzgerald_2006}, the probability of getting an instantaneous intensity $I$ given $I_c$ and $I_s$ is governed by the Modified Rician (MR) distribution, defined as:
\begin{equation}
    \rho_{\rm MR}\left[ I | I_c, I_s \right] = \frac{1}{I_s}\exp{\left[ - \frac{I + I_c}{I_s} \right] I_0 \left[ \frac{2 \sqrt{I\,I_c}}{I_s} \right]}, 
    \label{equ:mr}
\end{equation}
where $I_0[x]$ denotes the zero-order modified Bessel function of the first kind. The parameter $I_c$ represents the intensity of the ``constant'' part of the diffraction pattern, i.e.~the PSF of a star without the atmosphere, while $I_s$ is the intensity of the seeing halo which manifests as a ``speckle'' pattern. 

Typically the AO system attempts to confine all the speckles into the constant diffraction pattern. A coronagraph can remove or transform this coherent portion out of the image plane. For all of the calculations in this paper we assume that the Strehl ratio, and by extension $I_c$ and $I_s$, remain constant. 

Speckle intensity is correlated temporally. In the limit of Kolmogorov atmospheric turbulence and frozen flow, the speckle decorrelation time can be thought of as the wind crossing time across the telescope pupil \citep{Macintosh_2005}. In reality, the turbulence is not Kolmogorovic and the AO loop and telescope vibrations can further complicate the speckle temporal power spectrum density (PSD) \citep{Stangalini_2016}. This may result in a faster speckle decorrelation time, or a temporal PSD described by multiple exponential timescales. For the purposes of this paper we characterize the speckle PSD as a simple exponential decay with a characteristic speckle lifetime of $\tau_s = 0.1$~s to roughly match empirical data in the near infrared \citep{fitzgerald_2006, Meeker2018, Goebel_2018}. The speckle lifetime can change drastically depending on the atmospheric conditions but our results here can be qualitatively understood in those cases by scaling all parameters in time. A core condition for the SSD technique is for many photons to arrive in a single speckle decorrelation time: if this is satisfied, individual speckle fluctuations can be probed. 

\vspace{20pt}   
\subsection{Modeling an Incoherent Source}

A Poisson source incoherent with the stellar light and parameterized by intensity $I_p$ can be injected and the relevant photons uniquely identified in simulation.  Any incoherent Poisson sources will be represented by this term, including binary companions, planetary companions, extended sources, dark current, and read noise. In the case that $I_s=0$, the MR distribution reduces to a Poisson distribution with intensity $I_c$ and will be indistinguishable from $I_p$. In the demonstrations that follow, we identify $I_p$ as an injected planetary companion. 

The intensity fluctuations associated with speckles elevate the noise floor above the typical shot noise, i.e., it is harder to measure the planet's intensity when it is embedded in a boiling speckle field. If the speckle temporal information is marginalized over, then we show in the Appendix that the total variance of temporally correlated intensities obeying MR statistics combined with photon shot noise is 
\begin{equation}
    \sigma^2_{I, \rm tot} \approx \frac{2\tau_s \left( I_s^2 + 2 I_c I_s \right) + I_c + I_s + I_p}{T_{\rm tot}}.
    \label{eq:totalnoise}
\end{equation}
The variance of the measured intensity is inversely proportional to the total integration time, $T_{tot}$, as expected. This is the long exposure ($t_\mathrm{exp} \gg \tau_s$) photon noise limit faced by all PSF subtraction techniques like ADI and SDI. In the limit that $2\tau_s I_s \ll 1$, this variance reduces to pure shot noise on the number of photons. For high contrast imaging, typical parameters might be $\tau_s=0.1$~s, $I_s=50$~s$^{-1}$, and $2\tau_s I_s = 10$, in which case the noise from speckle fluctuations will dominate. In Section \ref{sec:binfree}, we show that this noise can be overcome by temporally resolving individual fluctuations in the speckle background. 

\subsection{Generating Mock Photon Lists}

We have developed code\footnote{Part of the MKID Pipeline python package available at https://github.com/MazinLab/MKIDPipeline} for quickly generating mock photon lists with an optional injected planet, corresponding to the output of a single MKID-like pixel which is single photon counting with low noise. The photon lists obey the following rules: 

\begin{enumerate}
    \item The underlying intensities are MR distributed, but are correlated in time with $\langle I(t)I(t + \delta t )  - \langle I \rangle^2 \rangle \propto \exp{\left[-\delta t / \tau_s\right]}$. 
    \item $I_c$, $I_s$, and $I_p$ are independently specified by the user, such that $I_c$ and $I_s$ govern the MR statistics of the stellar intensity and $I_p$ is the mean count rate of a Poisson source. The total intensity should have an expectation value of $\left\langle I \right\rangle = I_c + I_s + I_p$. 
    \item Due to the intrinsic dead time in an MKID, photons are removed from the list if they arrive within $\tau_0$ of the previous valid photon's arrival time. 
\end{enumerate}

The procedure for generating photon timestamps begins with creating a correlated list of random numbers that follow a Gaussian distribution. The random numbers are transformed to a uniform distribution ranging from 0 to 1, and finally transformed again to a MR distribution. The correlated MR sequence defines the intensity in photon counts for a small ($\ll \tau_s$) time bin (we used 200~$\mu$s), and a Poisson draw on that ``instantaneous'' intensity determines the number of photons that will finally be placed into that bin. The photons are distributed according to a uniform distribution in each bin. For simplicity, we assume neighboring pixels to have uncorrelated photon lists. 

Since MKIDs do not have the same dark current or readout noise as conventional semiconductor detectors we do not add any additional noise. However, during high count rates photons can be lost due to a firmware triggering lockout that acts as a non-paralyzable dead time \citep{Eyken_2015}. We use $\tau_0 = 10~\mu$s to match the latest firmware implemented on MEC. This dead time formulation could be used with quasi-photon counting EMCCDs to account for photon pile-up. 

\section{SSD with Millisecond Images} \label{sec:binned}

\begin{figure}
    \centering
    \includegraphics[width=\linewidth]{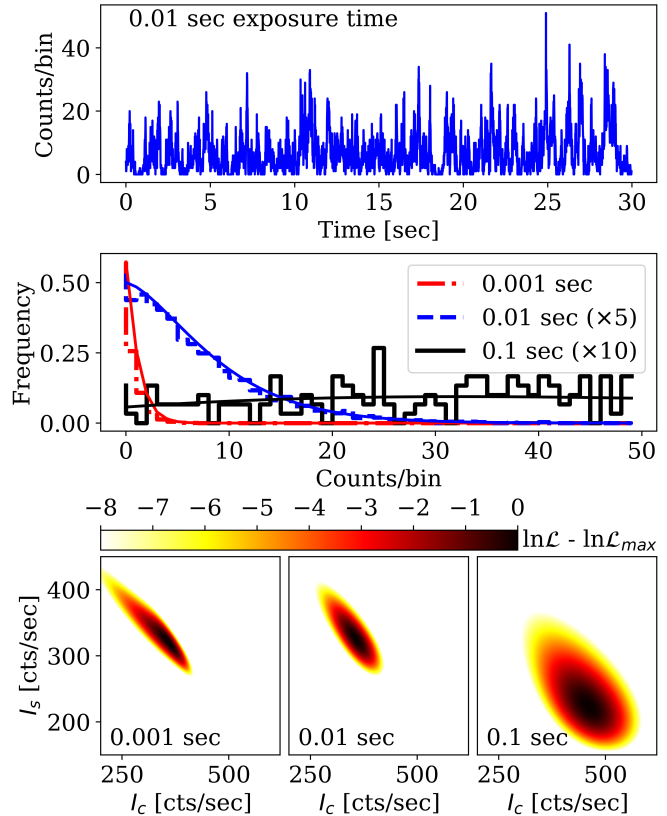}
    \caption{Top: a 30~s mock photon list simulated using the parameters $I_c = 300$, $I_s = 300$, $I_p = 0$ photons/s with $\tau_s=0.1$~s, is binned into 0.01~s exposures to form a light curve. Middle: The same photon list is binned by three different exposure times and used to plot intensity histograms. The best-fit modified Rician functions are overplotted. The distribution changes shape as the bin size is varied; the corresponding fitted parameters also change. The histograms for 0.01 and 0.1~s exposure times have been scaled by factors of 5 and 10 respectively. Bottom: the likelihood is marginalized over $I_p$ for the three different exposure times to illustrate how the best fit parameters evolve with bin size. The darkest area in the plot represents the maximum likelihood and determines the best fit values of $I_c$ and $I_s$. The same photon list was used for all plots. }
    \label{fig:SSD_bin}
\end{figure}

Past efforts have been successful in extracting $I_c$ and $I_s$ parameters of astrophysical sources from a series of millisecond images \citep{fitzgerald_2006, Gladysz_2010, Meeker2018}. In these experiments images were acquired with a fixed exposure time and the intensities at spatial locations under test were extracted into a light curve like that shown in the top panel of Figure \ref{fig:SSD_bin}. The distribution of intensities formed a histogram to which a modified Rician function could be fit. The middle panel of Figure \ref{fig:SSD_bin} shows this process for three exposure times. Since faint companions can masquerade as static speckles, a large $I_c/I_s$ ratio can be used as a merit function for planet detection in comparison to the rest of the image field \citep{Gladysz_2010,Meeker2018}. 

\subsection{Maximum Likelihood Model for Discrete Light Curves}

In this section we extend previous work to a more formal approach by finding the best-fit values of $I_c$, $I_s$, and $I_p$ with a maximum likelihood algorithm operating on the light curve. This allows for a direct detection of the non-stellar intensity, $I_p$, in the form of a point source like a planet or an extended source like a protoplanetary disk. Furthermore, we can naturally introduce photon noise from a low-intensity photon counting regime. By allowing the MR distribution in Equation \eqref{equ:mr} to suffer a Poisson-Mandel transformation \citep{cagigal_1999, aime04b}, the discrete stellar intensity distribution becomes
\begin{align}
p_\star [ n| I_c,I_s] &=\int_{0}^{\infty} \frac{I^n}{n!}\exp\left[-I\right]\rho_{\rm MR}[I]dI \nonumber\\
&= \frac{1}{I_s + 1}\left( 1 + \frac{1}{I_s} \right)^{-n} \exp\left[ -\frac{I_c}{I_s} \right] \nonumber \\ & \quad \times L_n\left[ \frac{-I_c}{I_s^2 + I_s} \right] \exp\left[  \frac{I_c}{I_s^2 + I_s}  \right],
\end{align}
where $L_n$ is the $n^{\mathrm{th}}$ Laguerre polynomial. The units of intensity for this section are number of photons per exposure time, $t_\mathrm{exp}$. If a planet (or some other source that is incoherent with the star) exists in the field, the intensity distribution at that location will be the discrete convolution of $p_\star$ with the planet's independent probability distribution $p_p$. To simplify, we assume the incoherent source has a Poisson probability distribution, $p_p [m | I_p] = \exp\left[ -I_p \right] I_p^m/m! $ with average intensity $I_p$. The likelihood of the $i$th bin of a light curve containing $k$ photons given $I_c$, $I_s$, and $I_p$ is 
\begin{equation}
\mathcal{L}_i = \sum_{m=0}^{k_i} p_p[m|I_p]p_\star[k_i-m|I_c,I_s],
\end{equation}
which becomes
\begin{align}
\label{equ:binnedLogL}
\mathcal{L}_i  =  &\frac{1}{I_s + 1} \exp\left[ -I_p -\frac{I_c I_s}{I_s^2 + I_s}  \right] \nonumber \\ &\times \sum_{m=0}^{k_i}  \frac{I_p^m}{m!} \left( 1 + \frac{1}{I_s} \right)^{-(k_i-m)} \ L_{k_i-m}\left[ \frac{-I_c}{I_s^2 + I_s} \right].
\end{align}
The likelihood of the entire light curve is $ \mathcal{L} = \prod_i \mathcal{L}_i$. We use a Newton conjugate-gradient search to find the most likely values of $I_c$, $I_s$, and $I_p$. 

The maximum likelihood estimate of the intensity distribution is overplotted onto the light curve histograms for three different exposure times in the middle panel of Figure \ref{fig:SSD_bin}. The bottom panel shows the likelihood functions marginalized over $I_p$ for these three exposure times. 

\subsection{Performance of Millisecond Imaging SSD}

\begin{figure*}
    \centering
    \includegraphics[width=0.95\textwidth]{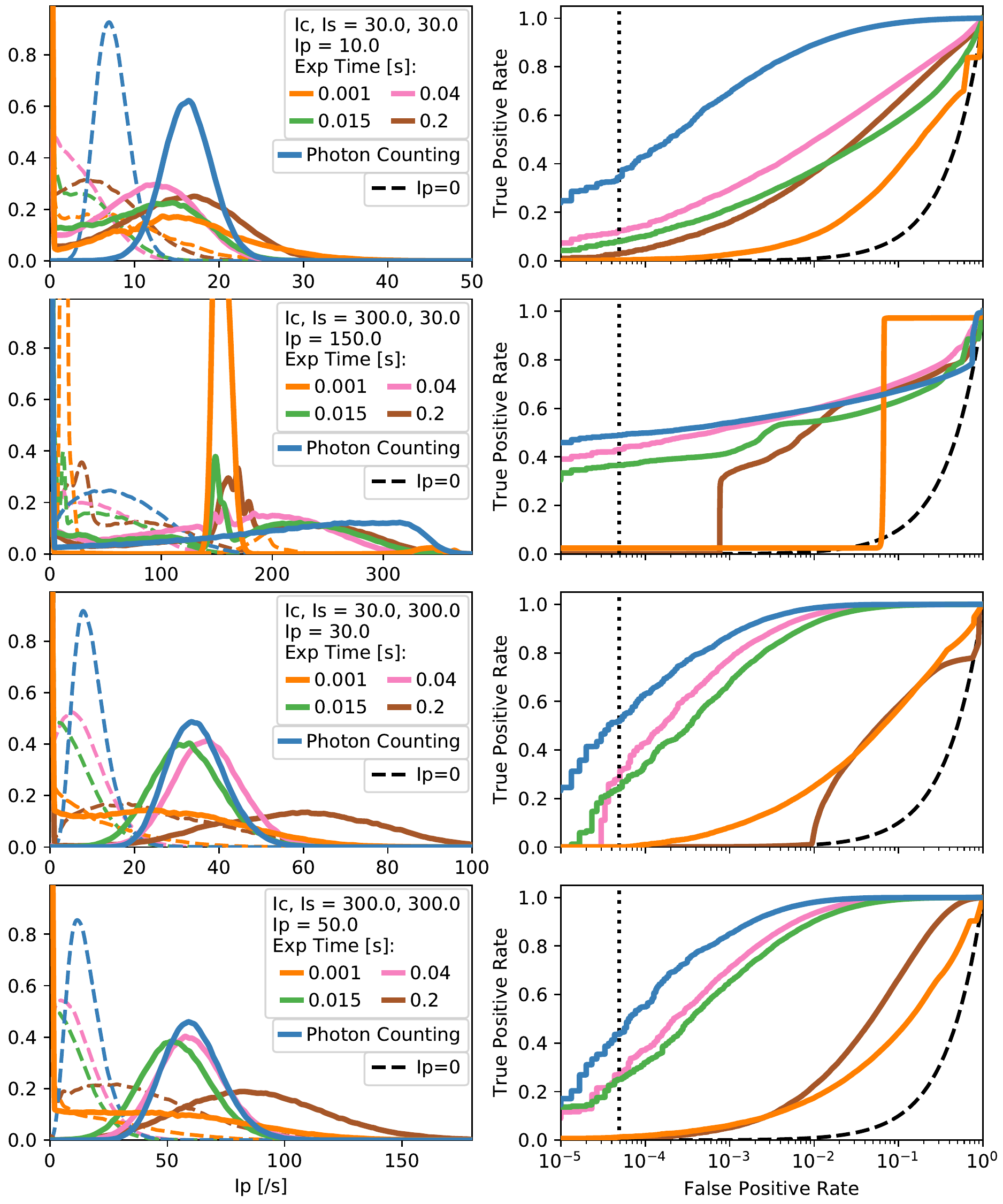}
\caption{Performance of our millisecond imaging SSD algorithm (Section \ref{sec:binned}) compared to our photon counting SSD algorithm (Section \ref{sec:binfree}). Left panels: histograms of the maximum likelihood estimates of $I_p$, computed using $3\cdot 10^5$ 30~s mock photon lists. Maximum likelihood estimates for $I_p$ are calculated for various effective exposure times and for the case with (solid line) and without ($I_p=0$, dashed line) an injected planet. The $y$-axis in the left column is arbitrary. The probability distributions are used to calculate the true/false positive rates for the receiver operator characteristic (ROC) curve (right). The vertical dotted line at 1/20000 (for the 20000 pixels in MEC) roughly indicates the maximum tolerable false positive rate. The full photon-counting SSD algorithm (blue lines) described in Section \ref{sec:binfree} outperforms the cases with nonzero exposure times.}
    \label{fig:rocVsBin}
\end{figure*}

To understand the performance of this millisecond imaging SSD algorithm and the extent to which it can improve exoplanet detections, we produce receiver operator characteristic (ROC) curves \citep{Tanner1954,DeLong+DeLong+Clarke-Pearson_1988,Krzanowski2009,Jensen_Clem_2017}. We achieve this by generating an ensemble of mock photon lists with the same nominal values of $I_c$, $I_s$, and $I_p=0$ and calculating maximum likelihood estimates to build up a distribution of $I_p$ corresponding to the signal-absent hypothesis. Next we inject a planet with $I_p>0$ and calculate maximum likelihood estimates to build a distribution for $I_p$ corresponding to the signal-present hypothesis. These distributions are shown in the left column of Figure \ref{fig:rocVsBin}. The cumulative distributions define the false positive and true positive fractions, which in turn define the shape of the ROC curve for a planet with a given intensity. 

The right column of Figure \ref{fig:rocVsBin} demonstrates the performance of the millisecond imaging SSD algorithm for different exposure times. The shape of the discrete intensity distribution can change depending on exposure time, which systematically affects the resulting maximum likelihood estimates (MLE) for $I_c$, $I_s$, and $I_p$ along with their uncertainties. In some cases there is a significant probability for the most likely value of $I_p$ to be equal to zero. In the case that $I_c \gg I_s$, the MLE probability can be multimodal with a significant peak at $\textrm{MLE}~I_p \approx \textrm{True}~I_c + \textrm{True}~I_p$. Both these behaviors appear for the same reason: when there is little modulation by $I_s$ of photons associated with $I_c$, then $I_c$ becomes difficult to distinguish from $I_p$. However, the total flux is still accurately recovered. 

\section{SSD in the Photon Counting Regime} \label{sec:binfree}

\subsection{Maximum Likelihood Model for Photon Arrival Times}

We find that there exists an optimal camera frame rate that leads to the most precise probability distribution for $I_c$, $I_s$, and $I_p$. For exposure times that are too long ($t_\mathrm{exp} \gg \tau_s$) the speckle temporal information is averaged over. For exposure times that are too short ($t_\mathrm{exp} \ll \delta t$, the photon inter-arrival time) the correlation between subsequent photon arrivals is lost because frames are interpreted as an unordered set containing only 0 or 1 photons. In order to avoid this pitfall, we develop the posterior probability for $I_c$, $I_s$, and $I_p$ directly from the set of photon inter-arrival times. 

We start by considering the (normalized) probability density for the next inter-photon arrival interval, $\delta t$, given a fixed intensity $I$:
\begin{equation}
\begin{aligned}
p[\delta t | I] = I e^{-I \delta t}.
\label{equ:poisson}
\end{aligned}
\end{equation}
The speckle field intensity is not fixed but varies in time with the modified Rician probability density described in Equation \eqref{equ:mr}. At a fixed point in time, the probability density of the next photon arrival time given $I_c$ and $I_s$ becomes
\begin{equation}
\begin{aligned}
p[\delta t | I_c, I_s] = \int_{0}^{\infty} p[\delta t | I] \rho_{\rm MR} [I | I_c, I_s] dI,
\end{aligned}
\end{equation}
where we have integrated the Poisson probability density from Equation \eqref{equ:poisson} over all possible instantaneous stellar intensities, $I$. Intensities are considered to be in units of photons per second. We assume here that the speckle lifetime is much longer than the time it takes a photon to arrive, $\tau_s \gg \delta t$. 

With a set of photon inter-arrival times, $\{\delta t_{i}\}$, we want the relative probability that a $\delta t$ is realized. At moments that happen to have higher instantaneous intensities the number of short $\delta t_{i}$'s will be increased. Thus, the relative probability of a photon inter-arrival time being in our data becomes
\begin{equation}
\begin{aligned}
p[\delta t | I_c, I_s] \propto \int_{0}^{\infty} p[\delta t | I] \rho_{\rm MR} [I | I_c, I_s] I dI.
\end{aligned}
\end{equation}
Finally, we consider the case that light incoherent with the star, such as from a planet, is in the field. We consider only the simplest case in which this source has constant intensity $I_p$ and is governed by Poisson statistics. In this case, the relative probability of $\delta t$ given $I_c$, $I_s$, $I_p$ is
\begin{equation}
\begin{aligned}
p[\delta t | I_c, I_s, I_p] \propto \int_{0}^{\infty} p[\delta t | I + I_p] \rho_{\rm MR} [I | I_c, I_s] (I + I_p) dI,
\label{eq:mr_deltat_like}
\end{aligned}
\end{equation}
which can be evaluated analytically. We then find the normalization constant by setting
\begin{equation}
\begin{aligned}
\int_{\tau_{0}}^{\infty} p[\tau | I_c, I_s, I_p] d\tau = 1.
\end{aligned}
\label{eq:normalization}
\end{equation}
Equation \eqref{eq:normalization} accounts for the non-paralyzable detector dead time, $\tau_0$, intrinsic to MKIDs by replacing the lower limit of integration with $\tau_0$. In Equation \eqref{eq:mr_deltat_like}, we set the likelihood equal to zero for $\delta t < \tau_0$. It is convenient to use the change of variables,
\begin{equation}
\begin{aligned}
u_i=\frac{1}{1+I_s \delta t_i}~,~~u_{max}=\frac{1}{1+I_s \tau_0}.
\end{aligned}
\end{equation}
The log-likelihood, $\log(\mathcal{L})[I_c, I_s, I_p]$, then becomes
\begin{align}\label{equ:binfreeLogL}
    \log(\mathcal{L})=& \sum_{i=1}^{N} p[\delta t_i | I_c, I_s, I_p] \nonumber\\
    =&\sum_{i=1}^{N}(u_i - 1) \left(\frac{I_p}{I_s u_i}+\frac{I_c}{I_s} \right)  \nonumber\\
    &+\sum_{i=1}^{N} \log\big[I_c^2 u_i^5 + 4 I_c I_s u_i^4 + (2 I_s^2 + 2 I_p I_c)u_i^3 \nonumber\\[-10pt]
    &~~~~~~~~~~~~~+ 2 I_p I_s u_i^2 + I_p^2 u_i\big]  \nonumber\\[3pt]
    &-N \frac{(u_{max}-1)(I_p + I_c u_{max})}{I_s u_{max}}  \nonumber\\[2pt]
    &-N \log\left[I_c u_{max}^3 + I_s u_{max}^2 + I_p u_{max}\right] .
\end{align}
As in Section \ref{sec:binned}, we assume that AO performance remains stable so that $I_c$ and $I_s$ remain constant over the course of observations. 

Equation \eqref{equ:binfreeLogL} for the photon counting SSD algorithm replaces Equation \eqref{equ:binnedLogL} from Section \ref{sec:binned}. We use a Newton conjugate-gradient search to find the maximum of the log-likelihood space and recover the best estimates for $I_c$, $I_s$, and $I_p$. 

\begin{figure*}
    \centering
    \includegraphics[width=0.95\textwidth]{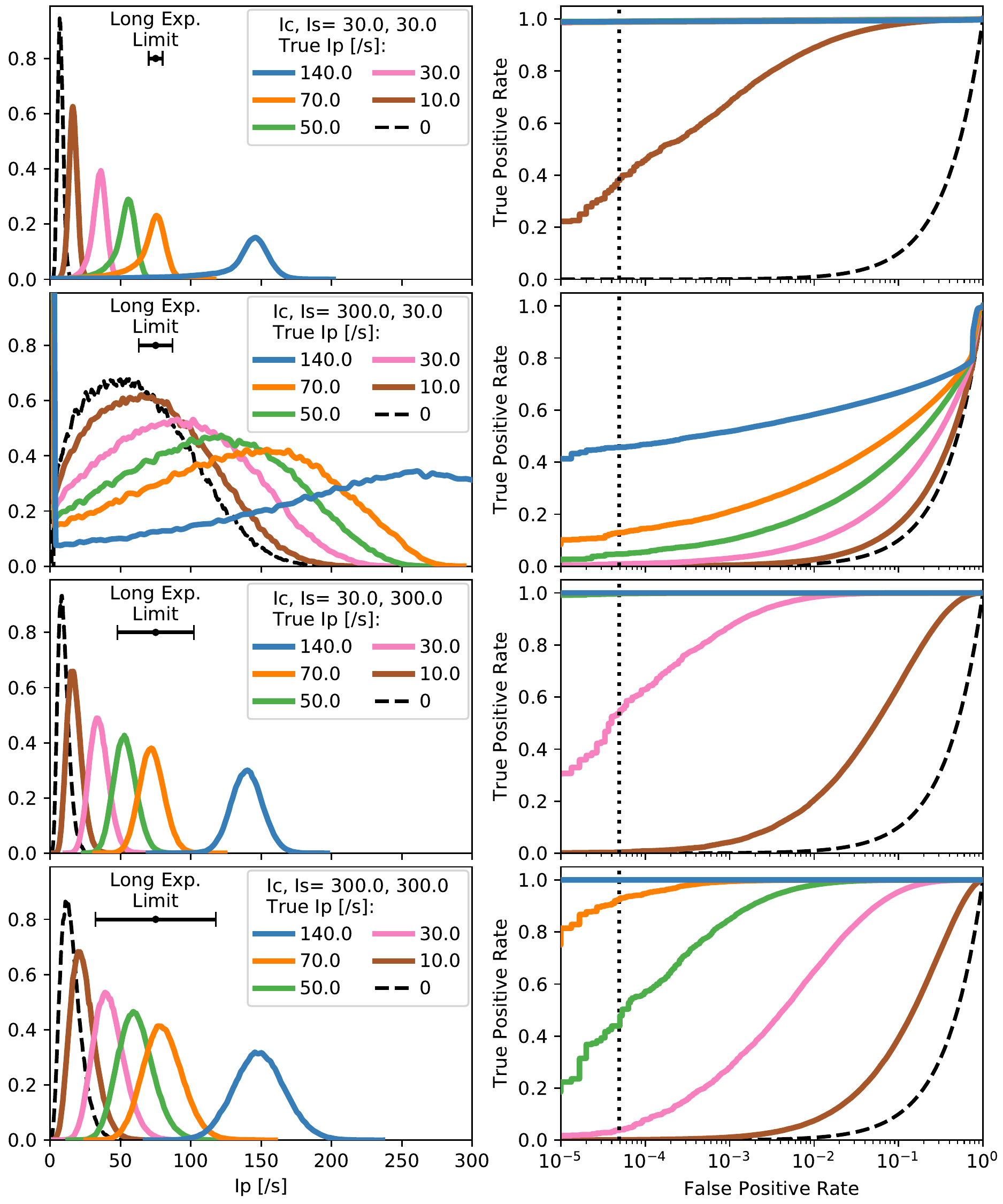}
    \caption{Performance of our photon-counting SSD algorithm. Left panels: histograms of the maximum likelihood estimates of $I_p$, computed using $3\cdot 10^5$ 30~s mock photon lists for each set of parameters. The $\pm\sigma$ for the long exposure photon noise limit (Equation \eqref{eq:totalnoise}) is shown with an error bar. The $y$-axis in the left column is arbitrary. The probability distributions are used to calculate the true/false positive rates for the receiver operator characteristic (ROC) curve (right panels). The vertical dotted line at 1/20000 (for the 20000 pixels in MEC) roughly indicates the maximum tolerable false positive rate. The algorithm performs well with a high true positive detection rate even for the case when $I_c$, $I_s$ are both large. However, the performance suffers in the case of $I_c \gg I_s$ (row 2). }
    \label{fig:rocVsIp}
\end{figure*}

The photon-counting SSD algorithm consistently outperforms the millisecond imaging SSD algorithm from Section \ref{sec:binned}, which marginalizes over temporal information via the exposure time (see Figure \ref{fig:rocVsBin}). In Figure \ref{fig:rocVsIp} we show the performance of the photon counting SSD algorithm under conditions given by various combinations of $I_c$ and $I_s$, and with various planet brightnesses $I_p$. The algorithm performs well with a high true positive detection rate even for the case when $I_c$ and $I_s$ are both large. However, the performance suffers in the case of $I_c \gg I_s$. As in millisecond imaging SSD, $I_p$ and $I_c$ become indistinguishable in this limit. 

\subsection{Maximum A Posteriori Estimation}

Prior knowledge of a parameter can improve the estimates of $I_c$, $I_s$, and $I_p$. In our case, we commonly have information on the $I_c$ parameter which corresponds to the static or quasi-static speckle point spread function either from a telescope model or measured on a reference star. For such situations, the log likelihood function in Equation \eqref{equ:binfreeLogL} can be modified with a Gaussian prior as
\begin{equation}
\begin{aligned}
\log(\mathcal{L}) \rightarrow \log(\mathcal{L}) - \frac{1}{2} \left(\left(I_c - \tilde{I}_c\right)/\sigma[\tilde{I}_c]\right)^2
\end{aligned}
\end{equation}
where $\tilde{I}_c \pm \sigma [\tilde{I}_c]$ is the prior on $I_c$. The new estimates of $I_c$, $I_s$, and $I_p$ become maximum a posteriori (MAP) estimates. 

\subsection{Performance on Simulated Telescope Image}

\begin{figure*}
    \centering
    \includegraphics[width=0.92\textwidth]{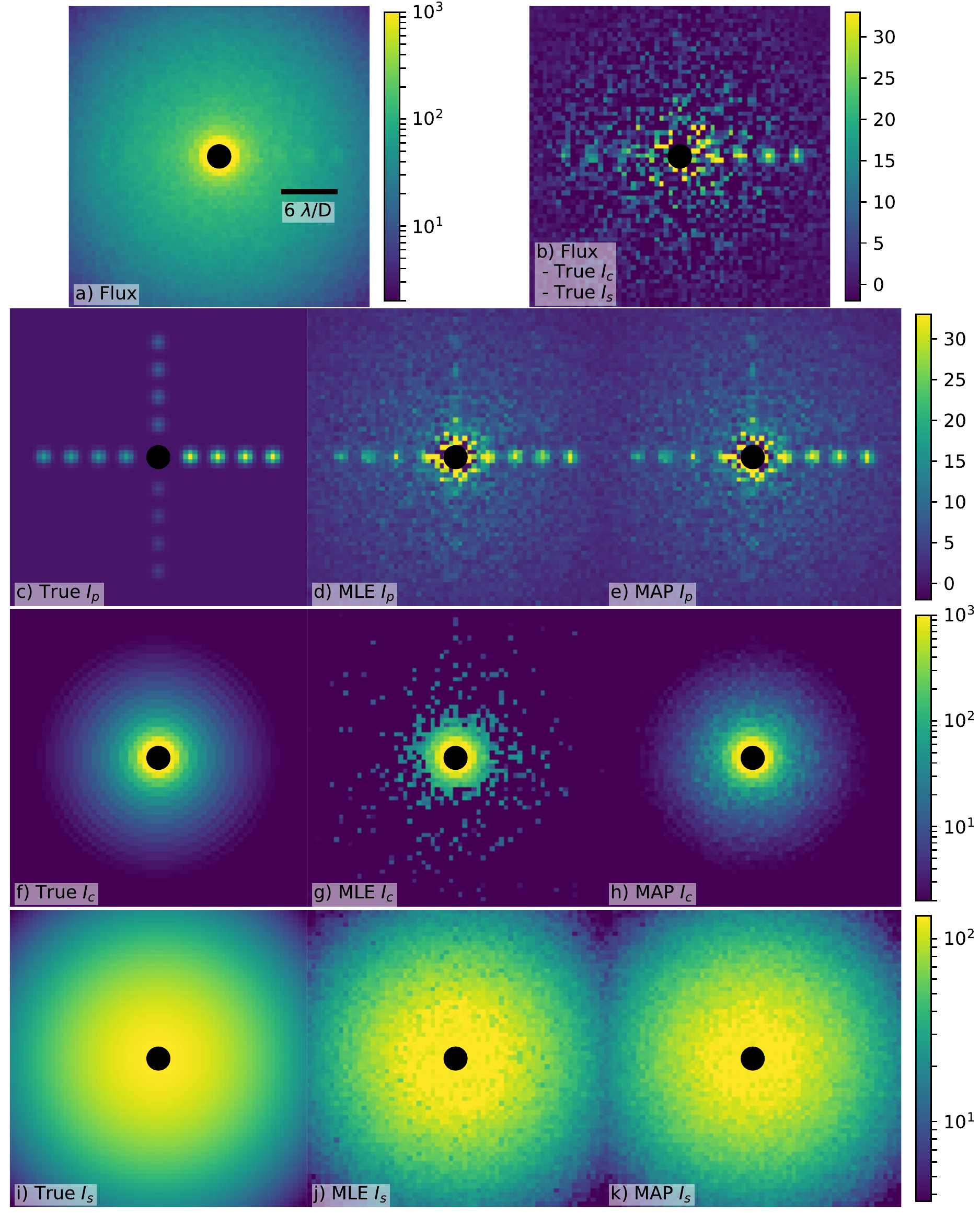}
    \caption{Performance of our photon-counting SSD algorithm on simulated telescope images.  Panel (a) shows the average intensity of simulated photon lists in each pixel; panels (c), (f), and (i) show the parameters used. Panel (b) subtracts stellar flux from (f) and (i) from the average intensity, (a), illustrating the speckle variance in a long exposure image. Panel (b) represents the theoretical limit of perfect PSF subtraction in a single exposure subject to MR intensity fluctuations. The photon-counting SSD algorithm results in the maximum likelihood estimates shown in (d), (g), and (j). Using a priori knowledge of the $I_c$ parameter (we used a Gaussian prior of True $I_c \pm 3\cdot \sqrt{\text{True}~I_c}$) we generate the maximum a posteriori (MAP) estimates shown in (e), (h), and (k). The planet signals extracted from the MAP $I_p$ estimate in (e) are not significantly improved compared to the MLE $I_p$ in (d). However, the SSD results in (d) and (e) both extract the planet better than the perfect PSF subtraction shown in (b). All images represent 30~seconds of data on a magnitude $J=10$ star with an 8.2~m telescope; all units are photons/second. 
    }
    \label{fig:image}
\end{figure*}

We evaluate the performance of the photon-counting SSD algorithm on a simulated 30~s telescope image without a coronagraph. We identify $I_c$ as the Airy-ring pattern of a diffraction limited telescope with a circular unobstructed aperature, $I_s$ as the seeing halo from atmospheric speckles with a Strehl ratio $\Sigma I_c / \Sigma (I_c + I_s) = 0.7$, and $I_p$ as a series of injected planets at various brightnesses and separations. The faintest planet's total intensity is 40~photons/s, for a contrast of $5 \cdot 10^{-5}$ with the host star, and is separated by 3.5~$\lambda$/D. Assuming a 5\% end-to-end throughput on an 8.2~m telescope, the stellar magnitude in the near infrared is approximately $J=10$. We assume a speckle decorrelation time of $\tau_s=0.1$~s and detector dead time of $\tau_0=10$~$\mu$s.

Each pixel has an independent 30~s photon list generated from the ``True'' $I_c$, $I_s$, and $I_p$ shown in panels (c), (f), and (i) respectively of Figure \ref{fig:image}. Spatial correlations in the photon lists are ignored for simplicity. The average intensity realized for each pixel is shown in Figure \ref{fig:image}a. Figure \ref{fig:image}b is the average intensity minus the expected light from the star, $(\textrm{True}~I_c + \textrm{True}~I_s)$, which illustrates the best possible long exposure PSF subtraction (compare the background variance to Equation \eqref{eq:totalnoise}). The MLE $I_c$, $I_s$, and $I_p$ are shown in Figure \ref{fig:image}d, g, and j respectively. In Figure \ref{fig:image}e, h, and k, we calculate the MAP estimates. We used the True $I_c \pm 3\cdot \sqrt{\text{True}~I_c}$ as a Gaussian prior on $I_c$; in practice one could use a telescope model or a reference PSF. The central black dot with radius 1.22~$\lambda$/D is not a coronagraph but simply obscures the on-axis light for convenience. 

Comparing Figure \ref{fig:image}d to Figure \ref{fig:image}b shows that the MLE $I_p$ from the photon counting SSD algorithm recovers the injected planets better than a perfect stellar PSF subtraction (i.e.~subtraction of the true $I_c$ and $I_s$). Figure \ref{fig:image}d and e show that the MAP estimate for $I_p$ is not significantly better than the MLE $I_p$ (although the MAP estimate for $I_c$ is more precise). This is surprising because the MAP estimate includes a prior on $I_c$ that should help discriminate between $I_c$ and $I_p$. That this is not the case indicates we can take full advantage of the photon counting SSD algorithm without prior knowledge of the telescope PSF. 

For the central pixel of each planet, we used $10^5$ independent photon lists to calculate the signal-to-noise ratio $\textrm{S/N}=(\langle I_p \rangle - \langle \textrm{Background}\rangle) / (\textrm{std. dev.} \langle I_p \rangle)$ where the $\langle \textrm{Background}\rangle$ is estimated by not injecting a planet. These are recorded in Table \ref{tab:mleSN}. The long exposure photon noise limit is also recorded in Table \ref{tab:mleSN} where the estimated $I_p$ is equal to the total flux minus the light from the star, $(\textrm{True}~I_c + \textrm{True}~I_s)$. While the results are from the simulated ensemble, they match the results from Equation \eqref{eq:totalnoise}. Table \ref{tab:mleSN} is representative of 30~seconds of data, but the S/N will scale with $\sqrt{T_{\rm tot}}$. For a 2 minute exposure, all values in Table \ref{tab:mleSN} should be scaled up by a factor of 2. The S/N was calculated using only the central pixel for convenience but would be larger if the surrounding pixels were considered.

\begin{deluxetable*}{lcccccccccccccccr}
\tablewidth{0pt}
\tablecaption{Companion SSD Signal-to-Noise Ratio}
\tablehead{ 
    \colhead{} &
    \colhead{} &
    \multicolumn{3}{c}{${\rm Separation} = 3.5\,\lambda/D$} &
    \colhead{} &
    \multicolumn{3}{c}{${\rm Separation} = 6.5\,\lambda/D$} &
    \colhead{} &
    \multicolumn{3}{c}{${\rm Separation} = 9.5\,\lambda/D$} &
    \colhead{} &
    \multicolumn{3}{c}{${\rm Separation} = 12.5\,\lambda/D$} \\
    \colhead{Contrast} &
    \colhead{} &
    \colhead{MLE} & \colhead{MAP} & \colhead{Limit\tablenotemark{$\dagger$}} &
    \colhead{} &
    \colhead{MLE} & \colhead{MAP} & \colhead{Limit\tablenotemark{$\dagger$}} &
    \colhead{} &
    \colhead{MLE} & \colhead{MAP} & \colhead{Limit\tablenotemark{$\dagger$}} &
    \colhead{} &
    \colhead{MLE} & \colhead{MAP} & \colhead{Limit\tablenotemark{$\dagger$}}
}
\startdata
        $4 \cdot 10^{-4}$ && 4.8 & 5.3 & 2.3 && 6.8 & 7.0 & 3.8 && 7.6 & 8.3 & 5.7 && 7.0 & 9.4 & 8.8 \\
        $2 \cdot 10^{-4}$ && 2.8 & 3.0 & 1.2 && 4.0 & 4.0 & 1.9 && 4.9 & 4.9 & 2.9 && 5.6 & 5.9 & 4.5 \\
        $1 \cdot 10^{-4}$ && 1.6 & 1.7 & 0.6 && 2.3 & 2.3 & 1.0 && 2.8 & 2.8 & 1.4 && 3.5 & 3.5 & 2.3 \\
        $5 \cdot 10^{-5}$ && 0.9 & 0.9 & 0.3 && 1.3 & 1.2 & 0.5 && 1.6 & 1.5 & 0.7 && 2.0 & 2.0 & 1.1
\enddata
\tablenotetext{$\dagger$}{Detection limit for a 30~s long exposure with perfect PSF subtraction of the stellar light.}
\label{tab:mleSN}
\end{deluxetable*}

\section{Discussion} \label{sec:discuss}

While it remains impossible to beat the photon shot noise $\sqrt{N}$, Figure \ref{fig:image} shows that the photon counting SSD algorithm can beat the long exposure ($t_\mathrm{exp} \gg \tau_s$) photon noise limit described by Equation \eqref{eq:totalnoise}. Table \ref{tab:mleSN} quantifies the improvement in the S/N as a factor of 3 in the case of faintest planet ($5\cdot 10^{-5}$) at the nearest separation ($3.5~\lambda/D$). This is possible because the speckle fluctuations are temporally resolved ($t_\mathrm{exp} \ll \tau_s$) and individual speckles are probed by multiple photons ($\delta t \ll \tau_s$). In the case that the fluctuations from stellar speckles dominate the variance of the total intensity ($2 I_s \tau_s \gg 1$), fast, noiseless detectors like MKIDs or EMCCDs are needed to dig beneath the noise. This often occurs in high contrast imaging at small separations ($\lesssim$ seeing radius) and is especially important at $\lesssim 5~\lambda/D$ where ADI and SDI start to lose their effectiveness (depending on spectral coverage, sky rotation, and AO performance). 

SSD will benefit ADI and SDI by reducing speckle noise from the data that is fed into those algorithms. ADI processing can be approached the same way as usual, but instead of using raw images one would use the $I_p$ images produced with SSD. The modulation of the planet location will be unaffected by SSD. Similarly for SDI, the algorithm would be given $I_p$ maps at various wavelengths. This approach would require wavelength information for each detected photon, which is an intrinsic feature of an MKID detector. 

Our SSD algorithm does not perform well when $I_c \gg I_s$ as seen in Figure \ref{fig:rocVsIp}. In this regime there is little modulation of the static speckle intensity $I_c$ by the atmospheric speckle field $I_s$, and as a result $I_c$ starts to become indistinguishable from the Poisson distributed $I_p$. This can be greatly mitigated with a coronagraph and possibly active speckle nulling \citep{Martinache_2014} which directly reduce $I_c$ in the image. 

On a space-borne telescope atmospheric speckles are not a concern, implying that the image-plane intensity will have different temporal behavior from ground-based observatories. While our SSD algorithm relies on the intensity following a MR distribution, in general any distribution can be used so long as it is known. It may also be possible to modulate speckles in a controlled way using onboard AO.

\section{Conclusions} \label{sec:conclusions}

In this paper we exploit photon arrival time statistics with a stochastic speckle discrimination (SSD) algorithm to distinguish planets from speckles. We first extend previous work with a formalized maximum likelihood algorithm operating on light curves with fixed, albeit fast, exposure times. We find that the likelihood space can sometimes result in bimodal behavior in the case of $I_c \gg I_s$. Additionally, the choice in exposure time can systematically skew the MLE of $I_p$ and inflate its variance. More generally, with a fixed exposure time, the performance will change as a function of parameters $I_c$, and $I_s$. This is a problem because $I_c$ and $I_s$ can change with observing conditions as well as with separation from the host star.

To overcome these difficulties, we have developed a new photon-counting SSD algorithm that calculates the maximum likelihood for $I_p$ directly from the individual photon arrival times. With this approach the likelihood space becomes smooth and unimodal and the precision is maximized. The planet detection performance can be better by a factor of 2 than perfect stellar PSF subtraction of a long exposure. This requires fast noiseless detectors like MKIDs. 

We have made several simplifying assumptions in our analysis. We take the speckle temporal PSD to be described by the single exponential timescale $\tau_s$, we assume $I_c$, $I_s$, and $I_p$ to remain constant, and we assume that the MR distribution accurately describes the off-axis stellar intensity. Finally, we ignore chromaticity. These assumptions represent avenues of exploration for future work. The speckle temporal PSD can be measured and used to more accurately simulate photon lists. Since the instantaneous Strehl can be measured, a future implementation of this algorithm might include that as a priori information in the log likelihood model. This would also inform additional variance on $I_p$ apart from the Poisson noise. 

SSD algorithms are ultimately constrained on two fronts. First, the photon arrival time $\delta t$ must be much shorter than the speckle decorrelation time $\tau_s$: we need many photons to characterize the properties of a materialized speckle. Second, the performance degrades when $I_c$ is large but $I_s$ is small, because the incoherent planet light masquerades as the static speckles described by $I_c$. Fortunately, this can be improved with a coronagraph. 

In return, SSD algorithms are most useful when $I_s$ is large, which is often inescapable at small inner working angles. Furthermore, the results are not directly dependent on separation (although they are dependent on $I_c$ and $I_s$, which are larger at small separations). This makes SSD a powerful post-processing technique at small inner working angles where it can complement more established techniques like ADI and SDI. At high speckle intensities, SSD can even outperform the theoretical limits of a perfect implementation of ADI and SDI.

\acknowledgments

This work was supported by the National Science Foundation Grant 1710385 and NASA ROSES grant NNX15AG23G. We would like to thank Michael Fitzgerald (University of California, Los Angeles) for helpful comments during the early stages of this work. 

\bibliographystyle{apj_eprint}
\bibliography{bibliography}

\appendix
\label{appendix}
\renewcommand{\theequation}{A\arabic{equation}}

We derive here the noise of a long ($t_\mathrm{exp} \gg \tau_s$) exposure subject to speckle statistics.  This is not simply photon shot noise, but arises due to the fluctuations of the modified Rician itself.  We consider a modified Rician parametrized by $I_c$ and $I_s$ with an exponential decorrelation time $\tau_s$.  The basic statistics of the distribution are
\begin{equation}
    \langle I \rangle = I_c + I_s
\end{equation}
\begin{equation}
    \sigma^2_I = \langle I^2 \rangle - \langle I \rangle^2 = I_s^2 + 2 I_c I_s
\end{equation}
\begin{equation}
    \langle (I_i - \langle I \rangle) (I_j - \langle I \rangle)  \rangle = \sigma^2_I \exp \left[ -\frac{|t_i - t_j|}{\tau_s} \right] .
\end{equation}

We wish to compute the variance of the mean intensity measured over a finite time interval $T_{\rm tot}$, which we divide into $N$ subintervals, each of length $\delta t$.
\begin{equation}
    \overline{I} = \frac{1}{N} \sum_i I_i .
\end{equation}
This is 
\begin{align}
    \langle \overline{I}^2 \rangle - \langle \overline{I} \rangle^2 &= \frac{1}{N^2} \bigg \langle \sum_i \sum_j I_i I_j \bigg \rangle - \frac{1}{N^2} \bigg \langle \sum_i I_i \bigg \rangle^2 \\
    &= \frac{1}{N^2} \bigg \langle \sum_i \sum_j \left( I_i - \langle I \rangle \right) \left( I_j - \langle I \rangle \right) \bigg \rangle \\
    &= \frac{\sigma^2_I}{N^2} \sum_i \sum_j \exp \left[ - \frac{|t_i - t_j|}{\tau_s} \right].
\end{align}

Now we will set $t_i = i \delta t$.  For the third step below, we use $\delta t/\tau_s \ll 1$. More generally, we take $\delta t \rightarrow 0$, $N \rightarrow \infty$.
\begin{align}
    \frac{1}{N^2} \sum_i \sum_j \exp \left[ - \frac{|t_i - t_j|}{\tau_s} \right] &= \frac{1}{N^2} \sum_i \sum_j \exp \left[ - \frac{|i - j|\delta t}{\tau_s} \right] \\
    &= \frac{1}{N^2} \sum_{i=0}^{N-1} \left( \sum_{j=0}^{N-i-1} \exp \left[ -\frac{j\delta t}{\tau_s}\right] + \sum_{j=0}^{i} \exp \left[ -\frac{j\delta t}{\tau_s}\right] - 1\right)\\
    &= \frac{1}{N^2} \sum_{i=0}^{N-1} \left( \frac{\tau_s}{\delta t} \left(2 - \exp \left[ -\frac{(N - i) \delta t}{\tau_s} \right] - \exp \left[ -\frac{(i + 1) \delta t}{\tau_s} \right] \right) - 1 \right)\\
    &= \frac{2 \tau_s}{N \delta t}  - \frac{2 \tau_s}{N \delta t} \frac{1}{N} \left(\sum_{i=1}^N \exp \left[ -\frac{i \delta t}{\tau_s} \right] \right) - \frac{1}{N} \\
    &= \frac{2 \tau_s}{N \delta t}  - 2 \left(\frac{\tau_s}{N \delta t} \right)^2 \left(1 - \exp \left[ -\frac{(N + 1) \delta t}{\tau_s} \right] \right) + \frac{2\tau_s}{N^2 \delta t} - \frac{1}{N} . 
\end{align}
Taking $N \rightarrow \infty$, $\delta t \rightarrow 0$, and $N \delta t = T_{\rm tot}$, we have
\begin{align}
    \langle \overline{I}^2 \rangle - \langle \overline{I} \rangle^2 &= \left( \frac{2 \tau_s}{T_{\rm tot}}  - 2 \left(\frac{\tau_s}{T_{\rm tot}} \right)^2 \left(1 - \exp \left[ -\frac{T_{\rm tot}}{\tau_s} \right] \right) \right) \sigma^2_I \\
    &= \left( \frac{2 \tau_s}{T_{\rm tot}}  - 2 \left(\frac{\tau_s}{T_{\rm tot}} \right)^2 \left(1 - \exp \left[ -\frac{T_{\rm tot}}{\tau_s} \right] \right) \right) \left( I_s^2 + 2 I_c I_s \right).
    \label{eq:mr_phot_limit}
\end{align}
In the limit of a very short integration, $T_{\rm tot} \ll \tau_s$, the prefactor is unity (as expected).  For an integration time much longer than the decorrelation time, $T_{\rm tot} \gg \tau_s$ (as is more typical), Equation \eqref{eq:mr_phot_limit} simplifies to
\begin{align}
    \langle \overline{I}^2 \rangle - \langle \overline{I} \rangle^2 &\approx \left( \frac{2 \tau_s}{T_{\rm tot}} \right) \left( I_s^2 + 2 I_c I_s \right).
    \label{eq:mr_phot_limit_simple}
\end{align}
Assuming $2 I_s \tau_s \gg 1$ (the inter-photon arrival time from the speckle field is much shorter than the decorrelation time $\tau_s$), Equation \eqref{eq:mr_phot_limit} dominates over simple shot noise $\sigma^2 = (I_c + I_s)/T_{\rm tot}$.  Interestingly, setting the number of independent realizations of the modified Rician equal to $T_{\rm tot}/\tau_s$ would miss the factor of two in Equation \eqref{eq:mr_phot_limit_simple}.  Adding shot noise back in, assuming $T_{\rm tot} \gg \tau_s$, and including a component $I_p$ incoherent with $I_c$ and $I_s$, we have
\begin{equation}
    \sigma^2_{I, \rm tot} \approx \frac{2\tau_s \left( I_s^2 + 2 I_c I_s \right) + I_c + I_s + I_p}{T_{\rm tot}}.
\end{equation}

\end{document}